RESEARCH ARTICLE

Open Access

Open Data

Open Code

Open Peer-Review

# Variability in seeds' physicochemical characteristics, germination and seedling growth within and between two French Populus nigra L. populations

Marlène Lefebvre[1], Marc Villar[1], Nathalie Boizot[1], Armelle Delile[1], Benjamin Dimouro[2], Anne-Marie Lomelech[3], Caroline Teyssier[1]

[1] : INRAE, ONF, BioForA, F-45075 Orléans, France
[2] : INRAE, GBFor, F-45075 Orléans, France
[3] : University of Bordeaux, Centre de Génomique Fonctionnelle, Plateforme Protéome, Bordeaux, France

**ABSTRACT**

To improve understanding of the physiology, ecology and regeneration requirements of black poplar (Populus nigra L.), a severely endangered tree species in Europe, we analysed the biochemical composition of seeds from 20 families derived from open pollination of 20 trees in two contrasting environments in France, along the Drôme and Loire rivers. Significant between-family differences were detected. Total lipid contents differed qualitatively between seed of the two populations and were higher in Drôme population (214 ± 29.8 versus 172 ± 19.8 mg.g$^{-1}$ dw). Soluble sugars were less abundant in seeds of the Drôme population (78.8 ± 10.0 versus 104 ± 25.3 mg.g$^{-1}$ dw). The protein content (322 ± 74.3 mg.g$^{-1}$ dw) and quantity of reserve proteins did not differ between seeds of the two populations. Measurements in three consecutive years showed that seeds of the Loire population had significantly higher dry weight (0.69 ± 0.10 versus 0.45 ± 0.10 mg.g$^{-1}$ dw), but this did not significantly affect the germination rate or root growth of the seedlings measured 5 days after sowing. However, one group of Loire seedlings had substantially longer juvenile roots than another group.
Overall, the results suggest that intrinsic qualities of the seeds are not critical factors in early stages and in selection pressure of the species' life cycle and recruitment.

# Introduction

The European black poplar, *Populus nigra* L., is a dioecious, heliophilous species of the Salicaceae that inhabits the riparian softwood forest ecosystems on floodplains in a large area, ranging from western and southern Europe to west Asia and North Africa (http://www.euforgen.org/species/populus-nigra/). The species was identified by Gurnell and Petts (2006) as an ecosystem engineer that contributes to the shaping of fluvial systems, as its life history and ecology are closely related to river patterns and processes (Karrenberg et al., 2002; Corenblit et al., 2014).

Black poplar is one of the most endangered tree species in Europe, and there are two major threats to its genetic resources (Lefèvre et al., 2001). First, alteration of riparian ecosystems by human activities, including regulation of floods with hydraulic engineering, has impaired its regeneration and favoured succession transition from pioneer salicaceous stands to hardwood forests (e.g. Nilsson et al., 2005; Kondolf et al., 2007). Second, introgression from other *Populus spp*. poses threats, as a few cultivars can strongly contribute to its pollen and seed pools (Vanden-Broeck et al., 2005; Chenault et al., 2011).

*P. nigra* usually forms local populations by colonizing open areas with alluvial soils. Its sexual regeneration is essential for gene flow, recombination being required for maintenance or recombination required for maintenance or enhancement of its natural populations' genetic variation, disease resistance and adaptation to environmental changes (Barsoum et al., 2004; Fady et al., 2015). However, in disturbed fluvial systems its regeneration is facilitated by high capacity for vegetative propagation, as reviewed by Chenault et al. (2011). The species' seeds are disseminated through wind and water, rapidly lose viability (Gosling, 2007; Kim, 2018), and have highly specific environmental requirements for germination. Large quantities of seeds are produced, mainly in May (in temperate Europe), coinciding with post-flood periods when freshly deposited, moist sediments are available for colonization (Karrenberg et al., 2002). This relationship between its life cycle and hydrological conditions has been described as the "biogeomorphological life cycle of *Populus nigra*" by Corenblit et al. (2014). The first two months after germination are critical for establishment of the seedlings, which require high soil moisture contents, so their roots must be able to grow down as rapidly as the water table recedes, according to the recruitment box model of Mahoney and Rood (1998). The seedlings may face many life-threatening stresses or disturbances during their first year of growth, including drought in summer, and prolonged inundation, burial or uprooting in the following autumn and winter (Mahoney and Rood, 1998; Corenblit et al., 2014). Thus, successful regeneration does not occur every year, and there is a strong age structure in natural stands, reflecting the flooding history (Wintenberger et al., 2019).

Black poplar trees reach the reproductive stage when they are 8-10 years old. In early spring (March), male and female trees produce flowers clustered in pendulous catkins. The female catkin is composed of 38-53 flowers and the ovary contains 10-13 ovules (Zsuffa, 1974; Karrenberg and Suter, 2003; Vanden Broeck et al., 2012 and personal data). Black poplar and other *Populus* species are prolific producers of seeds ('seed rain'). A single mature female tree can produce thousands or even millions of seeds annually (Karrenberg and Suter, 2003).

Hence, *Populus* seed are small (Stanton and Villar, 1996). For example, Zsuffa (1974) and Karrenberg and Suter (2003) found that mean masses of *P. nigra* seeds collected were 0.69 (sites of unknown origin) and 0.80 ± 0.05 mg (from northern Italy), respectively. However, available data are limited, and Karrenberg et al. (2002) found that seed masses vary by at least twice within the Salicaceae. The timing and duration of the seed release period of *Populus* (and other) species can also vary substantially, depending on environmental factors including photoperiod, atmospheric temperature and precipitation (Mahoney and Rood, 1998; Stella et al., 2006). For example, these phenological parameters may vary by up to 6-8 weeks in *Populus nigra* populations in temperate Europe.

Fresh *P. nigra* seeds are not dormant, germinate immediately (ca. 80-90%) according to several studies (Van Splunder et al., 1995; Foussadier, 1998; Guilloy-Froget et al., 2002; Karrenberg and Suter, 2003; González et al., 2016). However, the temperature during their maturation has been shown to influence their germination parameters in both controlled crossing experiments (Dewan et al., 2018) and natural conditions (Bourgeois and González, 2019).

Poplar seeds have been previously classified as recalcitrant (Gosling, 2007) since they have low dehydration tolerance, relatively high water contents at maturity, and high metabolic activity at dehiscence. Hence, only current-year seeds but no seed bank can germinate. For example, in a recent study, only four seeds (all of which were non-viable) were detected in surveys of 182 sampling spots (1 m x 1 m) in an alluvial bar in the middle Loire during August and September of four consecutive years (Greulich et al., 2019). The short viability of *Populus spp*. seeds has negative consequences for breeding and genetic conservation programs (Muller et al., 1982; Stanton and Villar, 1996). Hence, attempts have been recently made to address these problems and identify optimum protocols for ex situ seed conservation (Suszka et al., 2014; Michalak et al., 2015; Pawłowski et al., 2019). Following demonstrations that they can be stored up to several years in specific conditions, seeds of the species have been re-classified as intermediate between orthodox seeds, which can withstand dehydration to water contents as low as 50 mg.g$^{-1}$ (Roberts, 1973), and recalcitrant seeds (Bonner, 2008; Suszka et al., 2014; Michalak et al., 2015). Moreover, little information is available on the chemical constitution of seeds of tree species, including those widely used for wood production, and we have found no published information on the protein, carbohydrate and lipid compositions of *P. nigra* seeds.

The general idea of this study was to increase knowledge of the physiological and ecological characteristics of *P. nigra* seeds collected from natural populations. The two objectives were to characterize their morpho-biochemical composition and identify relationships (if any) between their composition and germination or early seedling growth at both intra- and inter-population levels. For this purpose, we collected seeds during three consecutive years from 10 trees in natural stands of each of two distinct populations. One population is located in floodplains of the middle Loire, in the northwest part of the Atlantic biogeographic region of France, and the other along the river Drôme (a tributary of the lower Rhône) in the Mediterranean region.

## Methods

**Plant material**
Seeds were collected from 10 trees in each of two genetically differentiated populations (Faivre-Rampant, 2016). One, designated the Drôme in the National Natural Park "Ramières du Val de Drôme". The other, designated the Loire population, is located on floodplains of middle reaches of the Loire river in (i) the National Natural Park of "Saint-Mesmin" and (ii) Guilly 40kms upstream. These sites are in areas of France with contrasting climates (Table 1). Hereafter, the seed lot collected from each tree is referred to as a 'family', and materials (trees, seeds or seedlings) collected or derived from the Drôme and Loire populations are respectively referred to as Drôme and Loire materials. The sampled trees in the Drôme and Loire populations were respectively designated D11-20 and L01-10. At maturity in May, capsules ready to split were collected, placed in paper bags tagged with a label showing the tree they were collected from and kept at room temperature. They were then separated from cotton and used fresh or frozen (at -80°C) until analysis. Capsules used for water content measurements were kept in closed plastic bags until measurement a few hours later.

**Table 1:** Geoclimatic data for the two sites. Data used to calculate the mean annual temperature were collected during 1971–2000 and provided by Météo-France. For 2018, data were collected from meteorological stations near sites of the populations. Temp.: temperature.

| Population | GPS coordinates | Mean annual temp. 2018 (°C) | Minimum temp. 2018 (°C) | Maximum temp. 2018 (°C) | Mean annual temp. 1971-2000 (°C) |
|---|---|---|---|---|---|
| Drôme | 44°44'34.0''N 4°56'37.89''E | 13.8 | -9.3 | 37.7 | 12.7 |
| Loire | 47°48'43.8''N 2°17'34.4''E and 47°53'9.7''N 1°50'49.2''E | 12.4 | -2.2 | 29.7 | 10.9 |

**Analyse the seed structure by light microscopy**

Freshly dissected seeds were observed under a light stereomicroscope (M125, LEICA, Wetzlar, Germany), and 45 images were taken with varying foci and maximum intensity to reconstruct 3D images.

**Fresh weight (fw), dry weight (dw), and water content**

In 2017, 2018 and 2019, three samples of 500 seeds were collected from each selected tree (Annex Table 1) to determine their dry weights (dw) after oven-drying at 70°C for 24 h (Teyssier *et al.*, 2011), which were doubled to obtain 1000-seed weights (P1000 values). In 2019, five samples per tree of fresh Loire seeds (between 33 to 50 seeds by sample) were immediately weighed to estimate their fresh weight (fw), then their water content was calculated as (fw-dw)/dw and expressed as g $H_2O.g^{-1}$ dw (Dronne *et al.*, 1997). The accuracy of the measurement depends on the time between harvesting and measurement. This time constraint did not allow this measurement to be carried out on the Drôme population.

**Determination of total protein content**

Total proteins were extracted, electrophoretically separated and quantified as described by Teyssier et al. (2014). Briefly, total proteins were extracted from five biological replicates of seeds (20-30 mg fw of frozen material harvested in May 2018) with 0.5 mL of lysis buffer (10 % (v/v) glycerol; 2 % (w/v) SDS; 5 % (v/v) β-mercapto-ethanol; 2 % (w/v) poly(vinyl) polypyrrolidone; 50 mM Tris pH 6.8). The samples were incubated for 5 min at 95 °C, and extracted twice. Protein concentrations in the extracts were assessed by Bradford assays with BSA (Bovine Serum Albumin) as a standard, and results were expressed as soluble protein content in $\mu g.g^{-1}$ dw. Protein profiles were obtained by separating proteins in the extracts by sodium dodecyl sulphate-polyacrylamide gel electrophoresis (SDS-PAGE) with a 12-20% acrylamide gradient, then staining with colloidal Coomassie Brilliant Blue G-250 (CBB-G).

**Protein identification by mass spectrometry**

Storage proteins, were identified and quantified by liquid chromatography-tandem mass spectrometry (LC-MS/MS) analysis of the crude extracts. Three biological replicates of each type of sample (Loire and Drôme seeds) were analysed following Gautier et al. (2018). Briefly, each protein sample was loaded onto an SDS-PAGE gel and digested with trypsin. The eluted peptide mixture was analysed using an Ultimate 3000 nanoLC system (Dionex, Amsterdam, The Netherlands) equipped with a C18 PepMap™ trap column (LC Packings) coupled to an Electrospray Q-Exactive quadrupole Orbitrap mass spectrometer (Thermo Fisher Scientific, San Jose, USA). Proteins were identified by SEQUEST searches, implemented via Proteome Discoverer 1.4 (Thermo Fisher Scientific Inc.), against a *P. trichocarpa* database from NCBI (NR_190603_Populus_trichocarpa.fasta, 68453 entries).

**Quantification of carbohydrates**

Soluble carbohydrates and starch were identified and/or quantified following Gautier et al. (2018 and 2019). Briefly, ethanolic supernatants of powdered extract from 20 mg dw samples of seeds harvested in May 2018) were purified using activated charcoal (Merck) and poly(vinylpolypyrrolidone) (PVPP, Sigma), dried, suspended in water and injected into a Chromaster HPLC system (VWR-Hitachi) equipped with a Rezex$^{TM}$ RPM-Monosaccharide Pb$^{+2}$ (8%) column (Phenomenex), then eluted with ultrapure H$_2$O at a flow rate of 0.6 mL.min$^{-1}$. Carbohydrates in the eluates were quantitatively detected with an ELSD 85 detector (VWR Hitachi) and the peak areas were electronically integrated using OpenLAB CDS EZChrom (Agilent). Carbohydrates were identified by co-elution with standards (Sigma), quantified from calibration curves and expressed in mg.g$^{-1}$ dw. Starch contents of the samples were determined, in glucose equivalents, by analysing amyloglucosidase hydrolysates of residual pellets of the extracts after soluble carbohydrate extraction (Gautier *et al.*, 2019). Each sample was assayed in triplicate, and means obtained from the samples± SD are presented.

**Determination of total lipid contents**

Total lipids were assayed by a colorimetric method involving use of sulpho-phospho-vanillin originally applied in animal analyses (Cheng *et al.*, 2011) with adaptation for seeds. Five replicates of each sample (25 mg dw of seeds harvested in May 2018) were crushed with 0.5 ml of 1:2 (v/v) chloroform:methanol using a stainless steel grinding ball. Mixtures of 98% sulphuric acid (v/v) and extracted samples or control (sun oil) were incubated for 2 min at 90°C before and after addition of vanillin reagent. The absorbance of each mixture at 520 nm was then measured using a Multiskan Spectrum spectrophotometer (Thermo Fisher Scientific) in a 96-well plate and used to calculate each sample's lipid content in mg.g$^{-1}$ dw**.**

**Germination trials and measurement of seedlings' root lengths**

For seedling analyses, seeds were stored at 3°C for at most 14 days after harvest (Annex Table 1), then sown by placing sets of 100 per family on wet filter paper in glass Petri dishes. After incubation for two days in a climate chamber (24 h at 25°C under day light), the number of germinated seeds was counted. This experiment was repeated with seeds collected from the same trees in 2017, 2018 and 2019. In addition, lengths of the roots of 38-55 seedlings per family were measured 5 days after the beginning of the trial, using a graduated ruler in 2019.

**Statistical analysis**

R version 3.6 (© 2009-2019 RStudio, Inc.) was used for all statistical analysis. For analyses of differences among families and between populations in measured variables of the seeds or seedlings' dw, protein, lipid and carbohydrate contents, and root length, we applied mixed models with lme4 and lmerTest packages (Bates *et al.*, 2014). This was done by considering genetic differences between families as random effects nested in fixed population (Loire or Drôme) effects. A mixed model was also used to assess effects of sampling year on P1000 values, with family as a random effect nested in population as a fixed effect and year (2017, 2018 or 2019) as a random effect. Data presented in x ± y format are means and SD. To explore and visualize correlations between families or populations and the measured variables, the data were subjected to Principal Component Analysis (PCA) with FactoMine R and Factoextra packages (Lê *et al.*, 2008).

# Results

**Seed structure and biochemical composition**

*P. nigra* seeds are small (less than 1 mm long), thin and develop in capsules grouped in catkins (Fig. 1). The average fw and dw were 2.11 ± 0.30 and 0.850 ± 0.10 mg, respectively (Table 2). Their water content at maturity was about 1.5 mg H$_2$O.mg$^{-1}$ dw, corresponding to 59.7% and 150% of their fresh and dry weights, respectively.

**Table 2:** Biochemical and physiological characteristics of seeds from the Drôme and Loire *P. nigra* populations. The seeds were harvested in 2018 if not otherwise indicated.

| | All sites | Drôme | Loire | Family effect | Population effect |
|---|---|---|---|---|---|
| dw | 0.569 ± 0.16 | 0.448 ± 0.10 | 0.689 ± 0.10 | *** | *** |
| dw (2019) | 0.850 ± 0.15 | n.d.[a] n.d. | 0.850 ± 0.15 | *** | *** |
| fw (2019) | 2.11 ± 0.31 | n.d. n.d. | 2.11 ± 0.31 | *** | [b] |
| WC (2019) | 1.50 ± 0.24 | n.d. n.d. | 1.50 ± 0.24 | *** | [b] |
| proteins | 322 ± 74.3 | 323 ± 83.0 | 320 ± 65.2 | *** | |
| lipid | 193 ± 32.8 | 214 ± 29.8 | 172 ± 19.8 | *** | *** |
| carbohydrates | 91.8 ± 23 | 78.8 ± 10.01 | 104 ± 25.3 | *** | ** |
| fructose | 0.456 ± 0.3 | 0.340 ± 0.22 | 0.573 ± 0.29 | *** | * |
| glucose | 1.44 ± 0.8 | 1.16 ± 0.50 | 1.72 ± 1.02 | *** | |
| myo inositol | 0.673 ± 0.2 | 0.635 ± 0.16 | 0.712 ± 0.28 | *** | |
| saccharose | 88.4 ± 22.8 | 76.1 ± 10.4 | 100 ± 25.3 | *** | ** |
| raffinose | 0.410 ± 0.2 | 0.509 ± 0.20 | 0.311 ± 0.11 | *** | ** |
| stachyose | 0.047 ± 0.1 | 0.000 ± 0.00 | 0.095 ± 0.20 | [c] | [c] |
| starch | 0.389 ± 0.17 | 0.299 ± 0.08 | 0.479 ± 0.19 | ** | *** |

Means obtained from analyses of 20 families (seeds collected from 10 trees at each site). By family, n = 3 for carbohydrate, n = 5 for total proteins and total lipids and n = 5 for dw (dry weight) determination. Values are given in mg.g$^{-1}$ dw ±SE, except for dw and fw (fresh weight) they are then in mg.seed$^{-1}$. Asterisks indicate significant effects (between-family and between-population differences) (***, P<0.001; **, P<0.01; *, P<0.05). [a] : not determined. [b] : not calculable as only Loire seeds were analyzed for this parameter. [c] : no statistical analysis possible since this sugar was only detected in two families. WC : water content.

Each seed consists of a thin, permeable and protective seed coat (testa) enclosing a yellowish green embryo composed of two cotyledons (which contain the nutrient reserves, as there is no endosperm), a hypocotyl and a radicle (Fig. 1B-E). The embryo accounts for almost all of the total seed weight. A few days after germination, the thick cotyledons provide the developing seedling with photosynthesis capacity (Fig. 1F), so they are classified as epigeal with reserve.

Protein, lipid and carbohydrate contents of the seeds accounted for ca. 44, 39 and 18% of their dry weight (Table 2). The protein profile was analysed from the crude protein extract of seeds. Its SDS-PAGE analysis resulted in bands indicative of more than 15 distinct major proteins, with masses ranging from 10 to 100 kDa (Fig. 2). These included some with electrophoretic mobility and apparent masses reminiscent of vicilin-like, legumin-like, and (most abundantly) albumin 2 storage proteins, the presence of which was confirmed by mass spectrometric analysis of the total protein extracts (Table 2). No visible differences in abundance of any storage proteins were observed between Drôme and Loire seeds (Table 3).

The seeds had high sucrose contents (88.4 mg.g$^{-1}$ dw), very low levels of several other soluble carbohydrates (<1.5 mg.g$^{-1}$ fw of glucose, myo-inositol, fructose, raffinose) and barely detectable levels of stachyose.

**Fig. 1:** Anatomical structure of *P. nigra* seeds at harvest time. **A** Global view of catkins releasing the cottonseeds. **B** Two full seeds out from the capsule, without the cotton. **C** Seed with seed coat partly removed. **D** and **E** Naked seed without seed coat. **F** 10-day-old seedling, with the two spatula-shaped cotyledons and two young leaves. Letters in the photos indicate the following structures: c, cotyledon; ca, catkin; co, cotton; cp, capsule; h, hypocotyl; r, radicle; sc, seed coat; se, seed. Scale bars represent 0.5 mm (B, C, D, E).

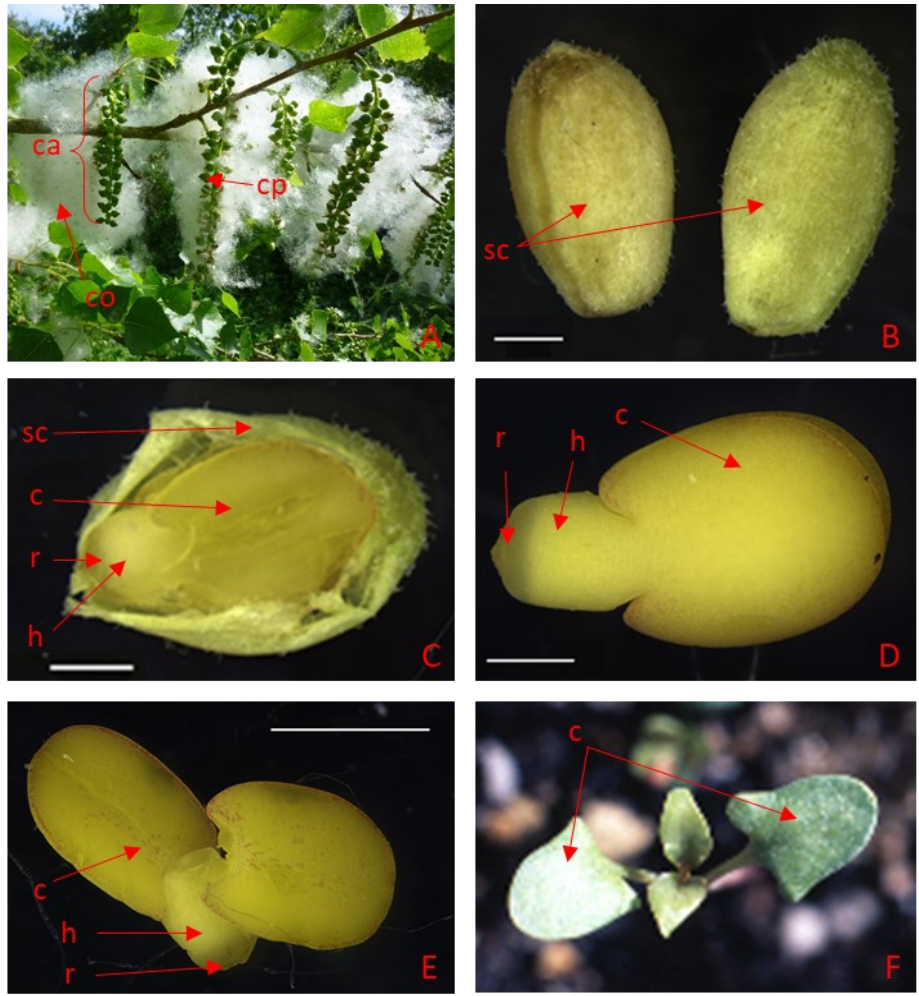

**Fig. 2:** Representative SDS-PAGE gel showing a total protein profile obtained from matured poplar seeds from tree families at the Drôme (lane D) or Loire (lane L) site. Molecular masses (kDa) of reference protein markers are indicated (lane MM). Arrows show bands corresponding to storage proteins identified by mass spectrometry.

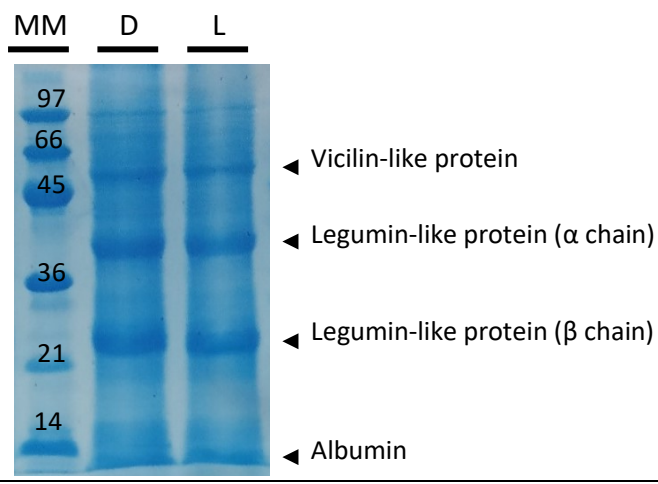

**Genetic diversity, at intra- and inter-population levels**

From 10 trees from each of two distinct populations, our data allow estimation of the variability of measured traits at both intra- and inter-population levels (Table 2), we detected larger ranges of carbohydrate contents of seeds per unit mass among Loire families (63.8-157 mg.g$^{-1}$) than among Drôme families (59.6-102 mg.g$^{-1}$), and the opposite pattern for their lipid contents (135-218 and 167-276 mg.g$^{-1}$, respectively). However, seeds of the two populations had large and similar variations in protein contents (Annex Fig. 1). Thus, population effects were detected in contents of carbohydrates (largely due to differences in fructose, saccharose and raffinose contents) and lipids, but not protein contents (Table 2). The first two Principal Components obtained from PCA accounted for 76.3% of the observed variance in these variables. P1000 was positively correlated with carbohydrate contents and both of these traits were negatively correlated with lipid contents. Protein content did not seem to be correlated with any of the other three traits (Fig. 3). The Loire seeds had higher P1000 values and carbohydrate contents, but lower lipid contents, than Drôme seeds. Elliptical envelopes of positions of the two populations in the PCA score plot overlapped, due to the similarity in their protein contents (Table 2) and large differences in scores between some families of the same population, especially the Drôme population.

**Table 3:** Identification by mass spectrometry of storage proteins in *P. nigra* seeds. The protein assignments and accession numbers were retrieved from the *P. trichocarpa* database (NCBI, *P. trichocarpa* (taxid:3694)). All descriptors were obtained from analyses of *P. trichocarpa* material.

| Description | Accession | % cov. | # AAs | MW [kDa] | Abund. Drôme | Abund. Loire |
| --- | --- | --- | --- | --- | --- | --- |
| 2S-albumin | PNT13296.1 | 18 | 148 | 17.1 | 2.29E+08 | 2.16E+08 |
| legumin A (11S-globulin) | PNT38121.2 | 44 | 527 | 59.3 | 3.92E+08 | 5.03E+08 |
| legumin B (11S-globulin) | PNS89766.1 | 49 | 487 | 55.1 | 3.09E+08 | 3.82E+08 |
| 11S globulin protein 2 | XP_002306851.3 | 36 | 512 | 58 | 2.24E+08 | 2.59E+08 |
| legumin B (11S-globulin) | XP_006370926.2 | 48 | 495 | 56 | 1.60E+08 | 1.80E+08 |
| legumin B (11S-globulin) | PNS89767.1 | 67 | 494 | 56 | 1.05E+08 | 1.01E+08 |
| vicilin-like peptides 2-1 | XP_006383667.2 | 36 | 597 | 67 | 1.81E+08 | 1.59E+08 |
| vicilin-like protein | XP_006380750.1 | 34 | 694 | 83 | 7.31E+07 | 7.03E+07 |
| vicilin-like protein | PNT19762.1 | 37 | 486 | 54.9 | 6.94E+07 | 6.78E+07 |
| bark storage protein A | ABK95364.1 | 11 | 339 | 36.4 | 6.36E+07 | 4.56E+07 |

% cov.: The percentage of the sequence covered by sequences identified in the included searches. MW [kDa]: Molecular weight of the protein. # AAs: Length of the protein sequence. Abund.: abundance (arbitrary unit).

**Variation among years in seeds' dry mass and germination traits**

P1000 values obtained for seeds collected in 2017, 2018 and 2019 varied among the three years ($\chi^2$=144, p<0.001), with a highly significant population effect (t=6.02, p<0.001) and family effect ($\chi^2$=128, p<0.001). Seeds of all Loire families were heavier in 2019 than in 2018 and 2017, but the patterns of between-year differences in P1000 values of the Drôme seeds were less clear and varied among families (Fig. 4). During the germination tests, the seedlings of different families were grown in Petri dishes under recommended cultural practices in order to determine the impact on growth among family. In germination tests, the germination rate varied between 0.85 and 1.0, depending on the family and year. Measurements of the roots of seedlings that developed from the germinated seeds 5 days after the beginning of the test in 2019 (with sets of 37-52 seedlings per family) revealed significant variations in their lengths within and between families ($\chi^2$=57.6, p<0.001), but not between the populations (t=-0.87, p>0.05, Fig. 5). Intriguingly, root lengths of seedlings of the D12, D13, L01, L02 and L05 families largely clustered in a single group, whereas seedlings of the other families formed two distinct groups: one with root lengths > 30 mm and the other with smaller (< 30 mm roots).

**Fig. 3:** Distributions of carbohydrate, lipid and protein components in the score plot for Principal Components 1 and 2 obtained from Principal Component Analysis and their contributions to explanation of the total variability of P1000 vales of the Drôme (closed circles and black ellipse) and Loire (triangles and grey ellipse) populations in 2018. n=3-5 per seed family depending on the biochemical parameter.

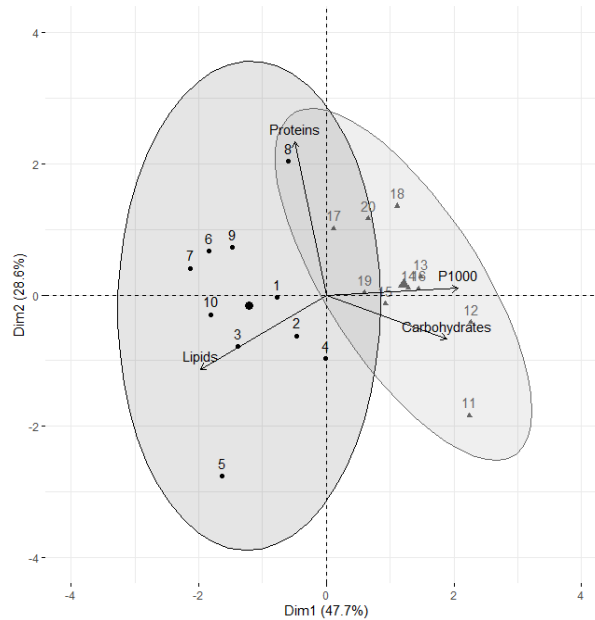

**Fig. 4:** Violin plots representing the density of observations of P1000 (mg) in each year (2017, 2018 and 2019) and population (Drôme in black circles and Loire in grey triangles). n=1-3 replicates per half-seed family. Each point or triangle represented a replicate.

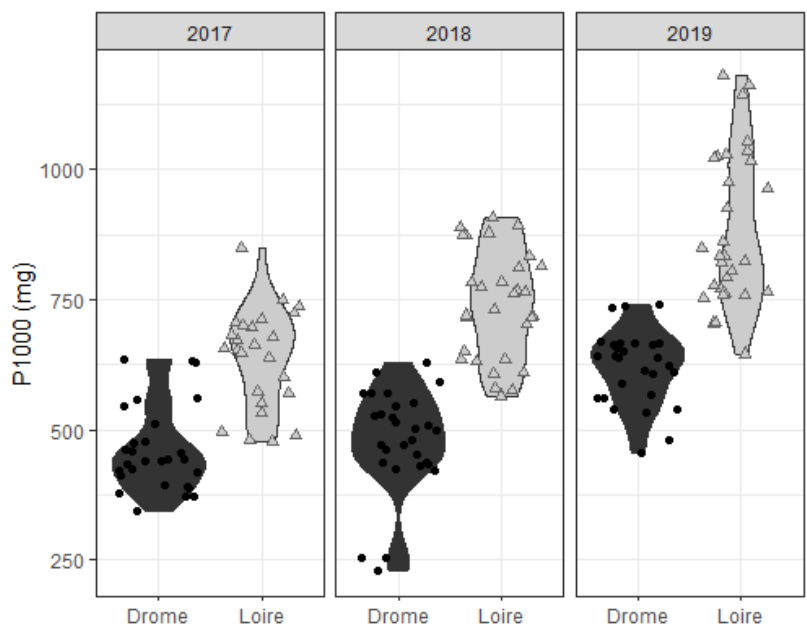

## Discussion

**Structural and biochemical characterisation of the seeds**

As already mentioned, *P. nigra* seeds have been classified as recalcitrant because of their rapid loss of viability under natural conditions (Gosling, 2007; Kim, 2018). Our observations of the following set of morphological and biochemical variables confirm this character. The testa surrounding the seed is very thin, and does not retard water loss from the seed (during the first 2 weeks following harvest, the seeds' water content dropped from about 60% to 15%, data not shown). They have a high water content at maturity. Their oligosaccharide content was very low. In addition, sucrose accounted for nearly all (96.6%) of the seeds' measured carbohydrates, and their raffinose and stachyose contents were low, in accordance with general reported traits of recalcitrant seeds (Steadman *et al.*, 1996; Lipavská and Konrádová, 2004; Egea *et al.*, 2009; Lipavská and Konrádová, 2004; Corbineau, 2012; Bishi *et al.*, 2013; Yada *et al.*, 2013; Wang *et al.*, 2018). These traits may play important roles in the species' reproductive cycle, as the seeds germinate on freshly deposited sandbars, which become available for colonisation at the end of Spring through recession of the water table. On the surface of these alluvial sediments the seeds must germinate very quickly after their release in May, otherwise they will have too little time to anchor strongly enough to survive the summer and autumn flood conditions (Mahoney and Rood, 1998; Corenblit *et al.*, 2014). Thus, rapid germination is essential for the seeds and the recalcitrant character of the *Populus* seeds is not a problem. Moreover, synthesis of complex sugars and other substances that could enable the seeds to withstand desiccation (Gösslova et al., 2001), would be a waste of energy. Conversely, high contents of sucrose may have high adaptive value as it can be rapidly converted into monosaccharides by invertases and used to meet metabolic as soon as germination is triggered. Following germination, a seed's energy reserves support the development of the seedling before it becomes autotrophic (Soriano *et al.*, 2013; Ghaffaripour *et al.*, 2017). In the absence of endosperm, the energy reserves are mainly stored in the cotyledons, but some may also be stored in the embryonic axis, which in some species may constitute a significant proportion of the seed mass. However, the embryonic axis accounts for a small proportion of the volume of a *P. nigra* seed (Fig. 1E). Thus, we chose to analyse energy reserves of whole seeds rather than specific tissues.

**Fig. 5:** Violin plots, representing the density of observations of lengths of roots of seeddlings of each family in each population (Drôme in black circles and Loire in grey triangles). n=37-52 seedlings per family. Each point or triangle represents a root length value.

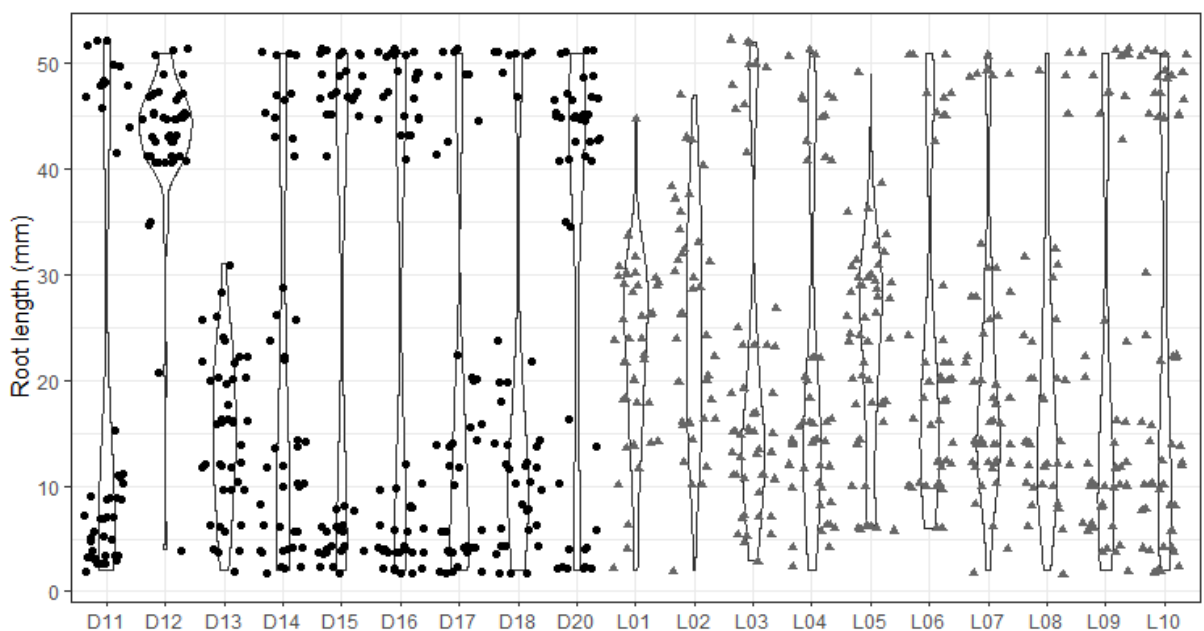

In plants, energy is stored in lipids, carbohydrates and reserve proteins, proportions of which vary greatly among species. Lipid and starch contents may be up to 50% (e.g. in rapeseed) to almost 80% (e.g. in wheat), respectively. We found that proteins, lipids and soluble carbohydrates all contribute substantially to energy reserves of *P. nigra* seeds (with ca. 53, 32 and about 15% contents, respectively). Their lipid contents approach those of oilseeds (e.g., 15-22, 25-50, and 30-50% of dw in soybean, sunflower, and rapeseed, respectively: Rodrigues *et al.*, 2012; Collective work, 2015).

Proteins in seed reserves play key roles, especially provision of seedlings with the amino acids required to synthesize new proteins. In accordance with their functional importance, the *P. nigra* seeds had far higher contents of storage proteins (ca. 60% of total proteins), than of structural and enzymatic proteins. Moreover, seeds of both populations had similar total protein contents and identical storage proteins, as identified by mass spectrometry. More specifically, levels of the main albumin, legumin- and vicilin-like storage proteins do not seem to differ significantly between seeds of the Drôme and Loire populations. These proteins are generally the most abundant proteins in dicotyledonous seeds. Vicilin- and legumin-like proteins belong to the globulin family (respectively 7S- or 11S-globulin) presenting a multigenic character (Shi *et al.*, 2010; Miernyk and Hajduch, 2011). This explains the presence of different isoforms identified by mass spectrometry (Table 3), and the presence of several gel bands (Fig. 2) correlated to these identifications (Teyssier *et al.*, 2014). The electrophoretic profile is very similar to those obtained in similar analyzes carried out on conifers (Morel *et al.*, 2014; Teyssier *et al.*, 2014) with a single band for the multimeric vicilin-like protein. The high presence of 2S-albumin (12.7% of energy storage proteins) is consistent with the recalcitrant nature of poplar seeds, due to its hydrophilic character inducing water retention capacity in tissues (Azarkovich, 2019). We also identified a "bark storage protein", representing 3.5% of the reserve proteins, which is also present in the xylem, cambium and parenchyma of poplar bark (Zhu and Coleman, 2001). In those tissues it reportedly stores carbon and nitrogen retrieved from senescing leaves, ready for use during vegetative growth in the following spring. It has been previously detected in immature poplar seeds (Zhu and Coleman, 2001), but its role in them has not been characterized. We hypothesize that this protein could be translocated into seedlings' stems, ready for use in the next leaf regeneration.

Lipid contents have been rarely assayed in non-food seeds, but at almost 20% dw in *P. nigra* seeds they are clearly important elements of the energy reserves. They are also substantially higher than the few previously reported contents in seeds of deciduous trees: 1.6 and 13.2% in *Castanea sativa* Mill. and *Quercus rubra L.* seeds, respectively (Akbulut *et al.*, 2017; Pritchard 1991). Lipids are less easily mobilized during germination than proteins or carbohydrates, but they have high potential importance as reserves because they store roughly twice as much energy per unit mass (Kitajima, 1996; Soriano *et al.*, 2011). Moreover, they may be the first energy reserves degraded in the processes that support seedlings' respiration and syntheses of fundamental metabolites before they establish full autotrophic capacity (Folkes, 1970; Soriano *et al.*, 2013).

**Difference in seed weight between Drôme and Loire**
In all three consecutive sampling years, Loire seeds had higher masses than Drôme seeds.
However, important variations in their mass were observed each year within each population. Several variables influence this mass (Castro *et al.*, 2006). We detected a genetic effect, an environmental effect and GxE interaction. Similar family effects have been previously found in other species such as oaks (González-Rodríguez *et al.*, 2012). Effects of environmental variables on seed masses have also been addressed, and temperature is the most important according to several previous studies (Soriano *et al.*, 2011; Lamarca *et al.*, 2013; Dewan *et al.*, 2018). This could at least partly explain our finding that Loire seeds had higher dw than Drôme seeds (Fig. 4), as the temperature is on average 1.8 °C warmer, according to 30-year mean data (Table 1), at the Drôme site than the Loire site. Many authors have also postulated that small seeds may be associated with fast ripening, and thus short filling phases (Fenner, 1992; Young *et al.*, 2004, Hampton *et al.*, 2013). Accordingly, our phenological data for the three study years indicate that on average there were only a difference of 2 days between flowering and seed dehiscence in the Drôme and Loire populations. The temperature of the maternal environment during development of *P. nigra* seeds may also significantly affect germination and seedling performance as suggested by Dewan et al. (2018). However, in contrast to sites of populations of other taxa compared in many previous studies, the difference in temperature between the Drôme and Loire environments is relatively small and does not seem to have any impact on seed quality. It would have been highly interesting to explore factors involved in between-year differences in seed weight, but further data are required. For example, soil variables govern nutrient bioavailability, which could influence seeds' masses (Castro *et al.*, 2006), but we have too little knowledge of differences in nutrient availability between the two sites to assess this possibility.

**Consequences of seed size for germination and growth of juvenile seedlings**
As the weight of a seed is linked to the energy reserves available for germination and early seedling growth, it seems reasonable to assume that a difference in mass should lead to a difference in vigour, and that seeds' sizes should correlate with survival rates. However, the relationships are highly taxon-dependent, as demonstrated by previously reported positive and negative correlations between these parameters (González-Rodríguez *et al.*, 2012; Rajjou *et al.*, 2012; Zhang *et al.*, 2015; Ghaffaripour *et al.*, 2017; Wang *et al.*, 2018). We observed no differences in three consecutive years in seedlings' germination rates, between either families or populations. Moreover, despite strong intra- and inter-family variations in seedlings' root growth, there was no significant difference in this variable between the two studied populations. The deviation from intuitive expectations of strong correlations could have been at least partly due to the optimized test conditions (Parker *et al.*, 2006). The high uncertainty in seeds' masses associated with their very small size (0.8 ± 0.15 mg) prevented individual-level assessment of correlation between seed mass and root elongation, but we observed strong bimodality in root lengths of some families' seedlings (D11, D14-20, L03, L04 and L06-L10; Fig. 5). Thus, we hypothesized that seedlings in the low-root growth groups would have minimal probabilities of survival if the water table rapidly receded (Mahoney and Rood, 1998), and that initial root growth could confer substantial selective advantage in such conditions.

Moreover, the dependence of seedlings on energy reserves depends on several parameters including the efficiency of the reserves' translocation into newly formed organs and the seedlings' photosynthetic efficiency (Soriano *et al.*, 2013), which depends on the type of cotyledon. We also found that rapid growth and acquisition of autotrophy via activation of photosynthesis in the cotyledons (which occurred within less than a day), as well as the conditions in our germination tests, may have contributed to the general absence of significant effects of seed weight and composition. These findings also corroborate the hypothesis that cotyledons' functional morphology is more strongly genetically conserved than seed weight (Zanne *et al.*, 2005). The growth of seedlings from light seeds such as poplar seeds is generally much more strongly dependent on photosynthesis establishment than on their contents of reserves (Zhang *et al.*, 2008). In addition, large seeds do not always fully utilize energy reserves during germination and seedling establishment (Kabeya and Sakai, 2003). For both of these reasons, seed mass is a poor predictor of germination success and seedling performance (González-Rodríguez *et al.*, 2012), and selective pressures favouring long-distance dissemination by wind may have driven *P. nigra*'s evolution of small seeds (Imbert and Lefevre, 2003).

**Conclusion**
The presented results provide the first description of the biochemical composition of *P. nigra* seeds, and both morphological and compositional confirmation of their recalcitrant character. Moreover, the data obtained regarding two populations in contrasting environments in France reveal significant variation in their composition (intra-family, inter-family within-population and between-population) as well as environmental effects. However, despite differences in mass in each of the study years between seeds from the Drôme and Loire populations, their germination parameters did not significantly differ, in accordance with the strong inter-family variability. These seed characteristics do not seem to be key factors for the successful recruitment and regeneration of *P. nigra*, and we hypothesise that the main selective pressures acting on the species' reproductive systems are the needs to coordinate flowering phenology and seed dispersal with local rivers' morphohydraulic dynamics and seed dispersal with the hydromorphological dynamics of the river.

# Data accessibility

Data are available online: https://doi.org/10.15454/H4P5U3

# Supplementary material

Script and codes are available online: https://doi.org/10.15454/H4P5U3

# Acknowledgements

We gratefully acknowledge the valuable contributions of Jérome Armand and Jean-Michel Faton (National Natural Park "Ramières du Val de Drôme"), Michel Chantereau and Damien Hémeray (National Natural Park "Saint Mesmin"), Vincent Lejeune and Rémy Gobin, the help in data collection and/or experiments provided by Christophe Borel and the BioForA and GBFor INRAE units, and the valuable assistance of Marie Pégard and Clément Cuello in statistical analysis and production of macrophotos, respectively. We also gratefully acknowledged the helpful comments provided by Drs Françoise Corbineau, Marie-Anne Lelu-Walter and Luc Paques.
"Version 3 of this preprint has been peer-reviewed and recommended by Peer Community In Forest and Wood Sciences (https://doi.org/10.24072/pci.forestwoodsci.100004)"

# Conflict of interest disclosure

The authors of this article declare that they have no financial conflict of interest with the content of this article.
Caroline Teyssier is one of the PCI Forest & Wood sciences recommenders.

# Appendix

**Annex Table 1:** Maturity (number of days between flowering and seed release) and date of harvesting seeds in each study year from 10 trees of both the Drôme population (D11-D21) and Loire population (L01-L10). [a]: not determined.

| Tree | Maturity 2017 | Maturity 2018 | Maturity 2019 | Harvest date, 2017 | Harvest date, 2018 | Harvest date, 2019 |
|---|---|---|---|---|---|---|
| D11 | 39 | 45 | 40 | 21-Apr | 25 Apr. | 26 Apr. |
| D12 | 52 | 54 | 51 | 04-May | 09-May | 9-May |
| D13 | 52 | 54 | 51 | 04-May | 09-May | 9-May |
| D14 | 52 | 54 | 51 | 04-May | 09-May | 9-May |
| D15 | 51 | 54 | 50 | 03-May | 09-May | 10-May |
| D16 | 51 | 54 | 50 | 03-May | 09-May | 10-May |
| D17 | 51 | 54 | 50 | 03-May | 09-May | 10-May |
| D18 | 52 | 54 | 51 | 04-May | 09-May | 9-May |
| D20 | 51 | 54 | 50 | 03-May | 09-May | 10-May |

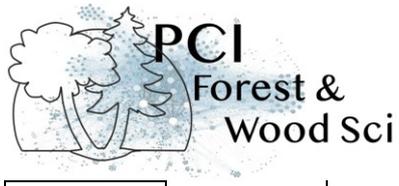

| | | | | | | |
|---|---|---|---|---|---|---|
| D21 | 52 | 54 | n.d. [a] | 04-May | 09-May | 9-May |
| L01 | 54 | 38 | 56 | 10-May | 13-May | 16-May |
| L02 | 61 | 44 | 62 | 30-May | 13-May | 22-May |
| L03 | 49 | 38 | 56 | 05-May | 13-May | 16-May |
| L04 | 60 | 44 | 56 | 10-May | 13-May | 16-May |
| L05 | 60 | 44 | 62 | 30-May | 13-May | 22-May |
| L06 | 59 | 42 | 62 | 10-May | 14-May | 22-May |
| L07 | 59 | 42 | 62 | 29-May | 14-May | 22-May |
| L08 | 59 | 42 | 62 | 29-May | 14-May | 22-May |
| L09 | 71 | 42 | 62 | 29-May | 14-May | 22-May |
| L10 | 60 | 44 | 56 | 30-May | 13-May | 16-May |

**Annex Fig. 1:** Biochemical composition (carbohydrates, lipids and proteins contents) of seeds of each family. Each black point (Drôme population) or grey triangle (Loire population) represents a replicate (3-5 replicates per family).

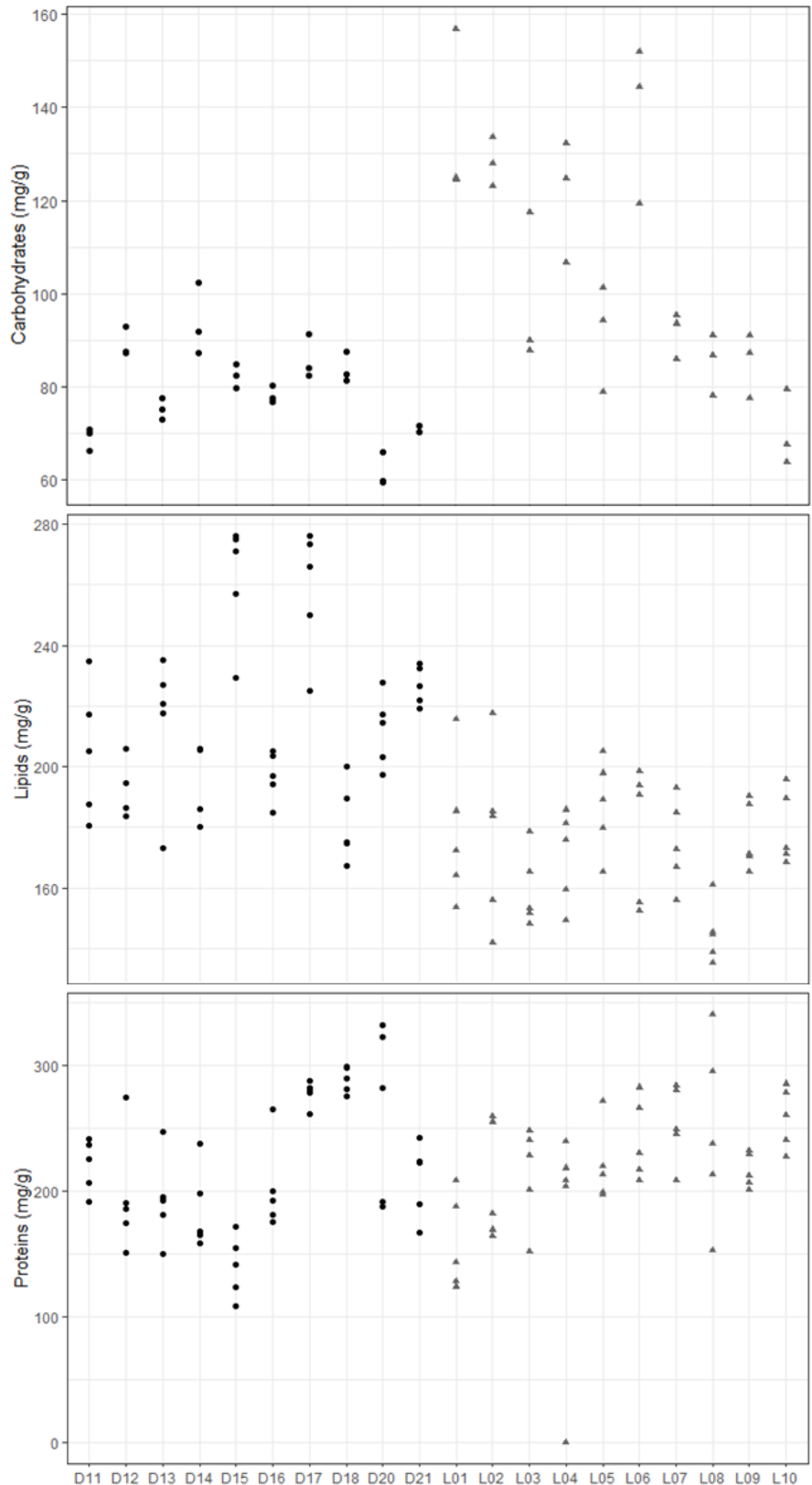

**Annex Table 2:** P1000 (mg) values obtained in 2017, 2018 and 2019 of *Populus nigra* family seeds from 20 trees of the Drôme population (D11-D21) and Loire population (L01-L10). n=3 per half-seed family.

| Genetic population | Tree ID | Replicate | Year | P1000 (mg) |
|---|---|---|---|---|
| Drôme | D12 | D12A | 2017 | 419.6 |
| Drôme | D12 | D12B | 2017 | 433.42 |
| Drôme | D12 | D12C | 2017 | 459.34 |
| Drôme | D13 | D13A | 2017 | 462.62 |
| Drôme | D13 | D13B | 2017 | 477.48 |
| Drôme | D13 | D13C | 2017 | 444.26 |
| Drôme | D14 | D14A | 2017 | 635.86 |
| Drôme | D14 | D14B | 2017 | 631.42 |
| Drôme | D14 | D14C | 2017 | 630.3 |
| Drôme | D15 | D15A | 2017 | 510.78 |
| Drôme | D15 | D15B | 2017 | 438.76 |
| Drôme | D15 | D15C | 2017 | 423.24 |
| Drôme | D16 | D16A | 2017 | 410.8 |
| Drôme | D16 | D16B | 2017 | 441.94 |
| Drôme | D16 | D16C | 2017 | 475.28 |
| Drôme | D17 | D17A | 2017 | 388.02 |
| Drôme | D17 | D17B | 2017 | 393.86 |
| Drôme | D17 | D17C | 2017 | 378.06 |
| Drôme | D18 | D18A | 2017 | 560.86 |
| Drôme | D18 | D18B | 2017 | 556.86 |
| Drôme | D18 | D18C | 2017 | 545.64 |
| Drôme | D19 | D19A | 2017 | 344.0 |
| Drôme | D20 | D20A | 2017 | 455.32 |
| Drôme | D20 | D20B | 2017 | 418.0 |
| Drôme | D20 | D20C | 2017 | 439.56 |
| Drôme | D21 | D21A | 2017 | 372.78 |
| Drôme | D21 | D21B | 2017 | 371.04 |
| Drôme | D21 | D21C | 2017 | 390.02 |
| Loire | L01 | L01A | 2017 | 706.00 |
| Loire | L01 | L01B | 2017 | 736.52 |
| Loire | L01 | L01C | 2017 | 711.8 |
| Loire | L02 | L02A | 2017 | 655.54 |
| Loire | L03 | L03A | 2017 | 663.02 |
| Loire | L03 | L03B | 2017 | 678.58 |
| Loire | L03 | L03C | 2017 | 724.46 |
| Loire | L04 | L04A | 2017 | 652.6 |
| Loire | L04 | L04B | 2017 | 672.06 |
| Loire | L04 | L04C | 2017 | 638.36 |
| Loire | L05 | L05A | 2017 | 682.0 |
| Loire | L05 | L05B | 2017 | 698.54 |
| Loire | L05 | L05C | 2017 | 700.56 |
| Loire | L06 | L06A | 2017 | 480.74 |
| Loire | L06 | L06B | 2017 | 495.5 |
| Loire | L06 | L06C | 2017 | 489.34 |
| Loire | L07 | L07A | 2017 | 477.04 |
| Loire | L08 | L08A | 2017 | 646.24 |
| Loire | L08 | L08B | 2017 | 749.22 |
| Loire | L08 | L08C | 2017 | 531.82 |
| Loire | L09 | L09A | 2017 | 600.76 |
| Loire | L10 | L10A | 2017 | 849.12 |
| Loire | L11 | L11A | 2017 | 570.68 |
| Loire | L11 | L11B | 2017 | 573.18 |
| Loire | L11 | L11C | 2017 | 550.3 |
| Drôme | D11 | D11A | 2018 | 627.9 |
| Drôme | D11 | D11B | 2018 | 610.88 |
| Drôme | D11 | D11C | 2018 | 593.14 |
| Drôme | D12 | D12A | 2018 | 452.18 |
| Drôme | D12 | D12B | 2018 | 462.94 |
| Drôme | D12 | D12C | 2018 | 460.96 |
| Drôme | D13 | D13A | 2018 | 424.04 |
| Drôme | D13 | D13B | 2018 | 436.18 |
| Drôme | D13 | D13C | 2018 | 421.26 |
| Drôme | D14 | D14A | 2018 | 570.88 |
| Drôme | D14 | D14B | 2018 | 571.5 |
| Drôme | D14 | D14C | 2018 | 569.86 |
| Drôme | D15 | D15A | 2018 | 514.52 |
| Drôme | D15 | D15B | 2018 | 508.6 |
| Drôme | D15 | D15C | 2018 | 497.9 |
| Drôme | D16 | D16A | 2018 | 253.3 |
| Drôme | D16 | D16B | 2018 | 253.8 |
| Drôme | D16 | D16C | 2018 | 227.8 |
| Drôme | D17 | D17A | 2018 | 480.12 |
| Drôme | D17 | D17B | 2018 | 471.38 |
| Drôme | D17 | D17C | 2018 | 471.32 |
| Drôme | D18 | D18A | 2018 | 501.06 |
| Drôme | D18 | D18B | 2018 | 526.84 |
| Drôme | D18 | D18C | 2018 | 522.08 |
| Drôme | D20 | D20A | 2018 | 529.02 |
| Drôme | D20 | D20B | 2018 | 551.12 |
| Drôme | D20 | D20C | 2018 | 546.32 |
| Drôme | D21 | D21A | 2018 | 434.06 |
| Drôme | D21 | D21B | 2018 | 431.66 |
| Drôme | D21 | D21C | 2018 | 435.6 |
| Loire | L01 | L01A | 2018 | 774.98 |
| Loire | L01 | L01B | 2018 | 768.72 |
| Loire | L01 | L01C | 2018 | 761.06 |
| Loire | L02 | L02A | 2018 | 878.86 |
| Loire | L02 | L02B | 2018 | 875.16 |
| Loire | L02 | L02C | 2018 | 875.24 |
| Loire | L03 | L03A | 2018 | 782.86 |
| Loire | L03 | L03B | 2018 | 765.02 |
| Loire | L03 | L03C | 2018 | 785.02 |
| Loire | L04 | L04A | 2018 | 650.432 |
| Loire | L04 | L04B | 2018 | 635.6 |
| Loire | L04 | L04C | 2018 | 630.76 |
| Loire | L05 | L05A | 2018 | 811.26 |
| Loire | L05 | L05B | 2018 | 815.42 |
| Loire | L05 | L05C | 2018 | 833.1 |
| Loire | L06 | L06A | 2018 | 609.02 |
| Loire | L06 | L06B | 2018 | 634.06 |
| Loire | L06 | L06C | 2018 | 608.58 |
| Loire | L07 | L07A | 2018 | 565.38 |
| Loire | L07 | L07B | 2018 | 577.5 |
| Loire | L07 | L07C | 2018 | 580.7 |
| Loire | L08 | L08A | 2018 | 718.5 |
| Loire | L08 | L08B | 2018 | 714.98 |
| Loire | L08 | L08C | 2018 | 730.94 |
| Loire | L09 | L09A | 2018 | 720.66 |
| Loire | L09 | L09B | 2018 | 704.22 |
| Loire | L09 | L09C | 2018 | 716.68 |
| Loire | L10 | L10A | 2018 | 908.14 |
| Loire | L10 | L10B | 2018 | 891.72 |
| Loire | L10 | L10C | 2018 | 890.42 |
| Drôme | D11 | D11A | 2019 | 640.86 |
| Drôme | D11 | D11B | 2019 | 621.7 |
| Drôme | D11 | D11C | 2019 | 612.9 |
| Drôme | D12 | D12A | 2019 | 538.56 |
| Drôme | D12 | D12B | 2019 | 533.76 |
| Drôme | D12 | D12C | 2019 | 538.34 |
| Drôme | D13 | D13A | 2019 | 662.54 |
| Drôme | D13 | D13B | 2019 | 665.98 |
| Drôme | D13 | D13C | 2019 | 669.7 |
| Drôme | D14 | D14A | 2019 | 736.72 |
| Drôme | D14 | D14B | 2019 | 740.02 |
| Drôme | D14 | D14C | 2019 | 734.5 |
| Drôme | D15 | D15A | 2019 | 588.26 |
| Drôme | D15 | D15B | 2019 | 566.18 |
| Drôme | D15 | D15C | 2019 | 561.76 |
| Drôme | D16 | D16A | 2019 | 478.88 |
| Drôme | D17 | D17A | 2019 | 662.42 |
| Drôme | D17 | D17B | 2019 | 658.38 |
| Drôme | D17 | D17C | 2019 | 640.92 |
| Drôme | D18 | D18A | 2019 | 667.26 |
| Drôme | D18 | D18B | 2019 | 637.46 |
| Drôme | D18 | D18C | 2019 | 667.44 |
| Drôme | D19 | D19A | 2019 | 650.04 |
| Drôme | D19 | D19B | 2019 | 608.6 |
| Drôme | D19 | D19C | 2019 | 639.22 |
| Drôme | D20 | D20A | 2019 | 608.96 |
| Drôme | D20 | D20B | 2019 | 561.7 |
| Drôme | D20 | D20C | 2019 | 455.1 |
| Loire | L01 | L01A | 2019 | 1145.2 |
| Loire | L01 | L01B | 2019 | 1182.02 |
| Loire | L01 | L01C | 2019 | 1163.9 |
| Loire | L02 | L02A | 2019 | 1029.14 |
| Loire | L02 | L02B | 2019 | 1036.4 |
| Loire | L02 | L02C | 2019 | 1053.3 |
| Loire | L03 | L03A | 2019 | 820.16 |
| Loire | L03 | L03B | 2019 | 824.68 |
| Loire | L03 | L03C | 2019 | 848.32 |
| Loire | L04 | L04A | 2019 | 770.4 |
| Loire | L04 | L04B | 2019 | 777.18 |
| Loire | L04 | L04C | 2019 | 767.0 |
| Loire | L05 | L05A | 2019 | 977.42 |
| Loire | L05 | L05B | 2019 | 965.5 |
| Loire | L05 | L05C | 2019 | 926.12 |
| Loire | L06 | L06A | 2019 | 793.82 |
| Loire | L06 | L06B | 2019 | 759.04 |
| Loire | L06 | L06C | 2019 | 804.94 |
| Loire | L07 | L07A | 2019 | 704.98 |
| Loire | L07 | L07B | 2019 | 645.98 |
| Loire | L07 | L07C | 2019 | 707.76 |
| Loire | L08 | L08A | 2019 | 861.04 |
| Loire | L08 | L08B | 2019 | 832.8 |
| Loire | L08 | L08C | 2019 | 833.9 |
| Loire | L09 | L09A | 2019 | 753.4 |
| Loire | L09 | L09B | 2019 | 763.82 |
| Loire | L09 | L09C | 2019 | 759.26 |
| Loire | L10 | L10A | 2019 | 1025.06 |
| Loire | L10 | L10B | 2019 | 1015.68 |
| Loire | L10 | L10C | 2019 | 1022.14 |